\documentclass{INTERSPEECH2023}

\interspeechcameraready 
\usepackage{multirow}
\usepackage{graphicx}
\usepackage[section]{placeins}
\usepackage{caption}
\usepackage{subcaption}
\usepackage[colorlinks=true, allcolors=blue]{hyperref}

\title{Data Augmentation for Diverse Voice Conversion in Noisy Environments}
\name{Avani Tanna, Michael Saxon, Amr El Abbadi, William Yang Wang}
\address{
  University of California, Santa Barbara}
\email{avani@ucsb.edu, saxon@ucsb.edu, amr@cs.ucsb.edu, william@cs.ucsb.edu}

\begin{document}

\maketitle
 
\begin{abstract}
Voice conversion (VC) models have demonstrated impressive few-shot conversion quality on the clean, native speech populations they're trained on. However, when source or target speech accents, background noise conditions, or microphone characteristics differ from training, quality voice conversion is not guaranteed. 
These problems are often left unexamined in VC research, giving rise to frustration in users trying to use pretrained VC models on their own data. 
We are interested in accent-preserving voice conversion for name pronunciation from self-recorded examples, a domain in which all three of the aforementioned conditions are present, and posit that demonstrating higher performance in this domain correlates with creating VC models that are more usable by otherwise frustrated users. 
We demonstrate that existing SOTA encoder-decoder VC models can be made robust to these variations and endowed with natural denoising capabilities using more diverse data and simple data augmentation techniques in pretraining.
\end{abstract}\noindent\textbf{Index Terms}: voice conversion, robustness, data augmentation

\section{Introduction}

Voice conversion (VC) is the task of generating utterances in a \textit{target speaker}'s voice that carry the content and prosody from a \textit{source utterance} from a different speaker \cite{661472}, preserving the content of the source utterance while reproducing the characteristics and style of the target speaker. 
VC was originally conceived of as a data-efficient way to add styles and personalities to 90s-era text-to-speech systems, to improve the quality of decoded speech in telephony, and as a way to preserve speaker individuality under speech translation \cite{MOULINES1995125}.
Many of these problems have been solved with other techniques, and VC has come to be treated as a novelty task for demonstrating innovations in data-efficient and few-shot generative modeling
\cite{qian2019autovc},
and real-world use-cases for VC technologies are no longer centered. 

However, niche real-world applications for VC still exist. 
One example is multicultural name pronunciation (e.g., in graduation ceremonies) where the desired standard is for one's own name to be read aloud as one pronounces it themself.
Typically, ceremony organizers solicit self-recordings of awardees voices, typically produced on their own cell phones. 
We propose treating these recordings as source utterances for conversion into the organizer's target voice. 
Under these conditions, represented phonemes differ considerably from those present in the VC model training speech distribution and utterances have inconsistent microphone characteristics and environmental noise. 
Unfortunately, we find that existing pretrained few-shot VC models, such as AutoVC \cite{qian2019autovc} and FragmentVC \cite{lin2021fragmentvc} perform poorly in these conditions.
Is it possible to adapt these SOTA VC models to be performant in this setting? Yes, we find.


\subsection{Related Work \& Existing Problems}

Encoder-decoder voice conversion models such as AdaIN-VC \cite{AdaIN-VC}, AutoVC \cite{qian2019autovc}, and AutoPST \cite{qian2021global} have recently set the standard for high-quality many-many voice conversion. 
A current state-of-the-art model in this vein, FragmentVC \cite{lin2021fragmentvc}, achieves high-quality parallel-data-free VC results using the pretrained Wav2Vec \cite{baevski2020wav2vec} speech encoder to characterize the source speech. 

It is challenging to document how to record a quality source utterance that will work well with a VC model produced on an unnatural distribution. 
We found that OOD sampling rate, volume levels, microphone quality, and clarity of speech due to distance all had significant impacts on output quality for the aforementioned models. 
This leads to significant frustration for users trying to use open-source VC models on their own data, evidenced by a litany of GitHub issues\footnote{\href{https://github.com/auspicious3000/autovc/issues/108}{github.com/auspicious3000/autovc/issues/108}, \href{https://github.com/auspicious3000/autovc/issues/28}{/28}, \href{https://github.com/auspicious3000/autovc/issues/19}{/19}, \href{https://github.com/auspicious3000/AutoPST/issues/14}{/14}}. 
Furthermore, these models are trained on speech from native English speakers (UK/US accents), which also was problematic for our target task. 
We chose FragmentVC \cite{lin2021fragmentvc} for adaptation because of its use of the pretrained Wav2Vec encoder, which we believed would be better able to handle the diverse set of source utterance phonemes.

\begin{figure}[t!]
    \centering
    \includegraphics[width=0.8\linewidth]{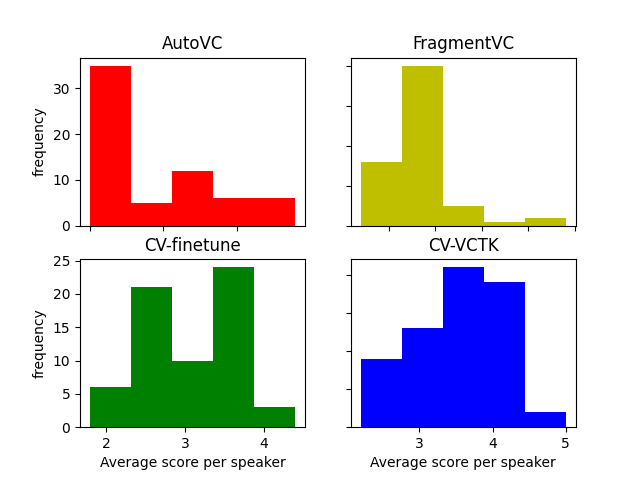}
    \caption{Speakerwise mean quality Likert score histograms for baselines and our robust models CV-finetune and CV-VCTK.}
    \label{fig:hist-avg-score-speaker}
    \vspace{-4ex}
\end{figure}

\subsection{Contributions}

\begin{figure}
     \centering
     \begin{subfigure}[b]{0.23\textwidth}
         \centering
         \includegraphics[width=\textwidth]{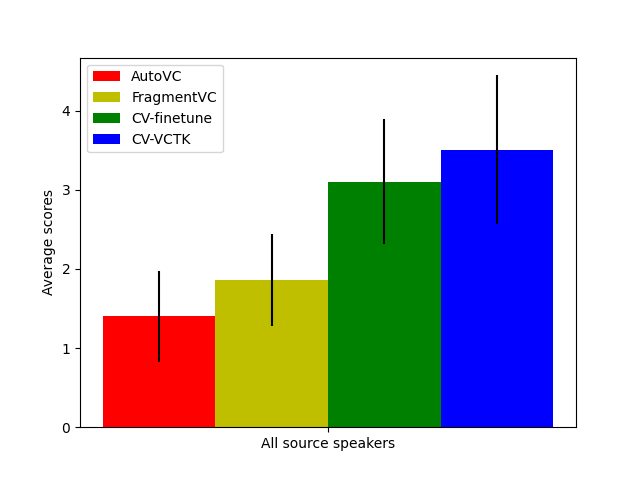}
         \caption{All speakers}
         \label{bar-avg-score-std-all}
     \end{subfigure}
     \hfill
     \begin{subfigure}[b]{0.23\textwidth}
         \centering
         \includegraphics[width=\textwidth]{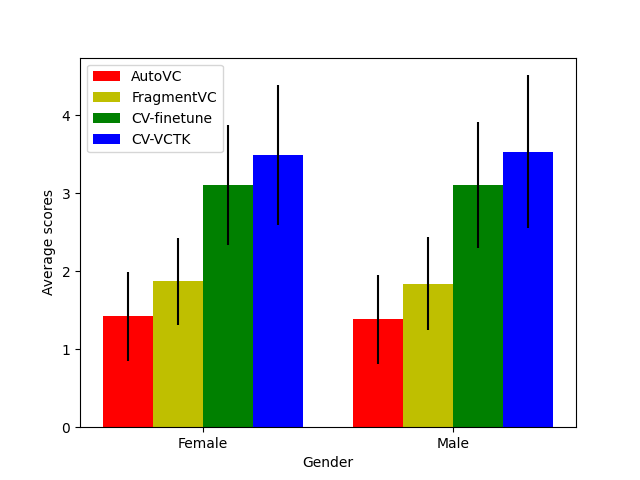}
         \caption{Speakers grouped by gender}
         \label{bar-avg-score-std-gender}
     \end{subfigure}
     \hfill
     \begin{subfigure}[b]{0.5\textwidth}
         \centering
         \includegraphics[width=\textwidth]{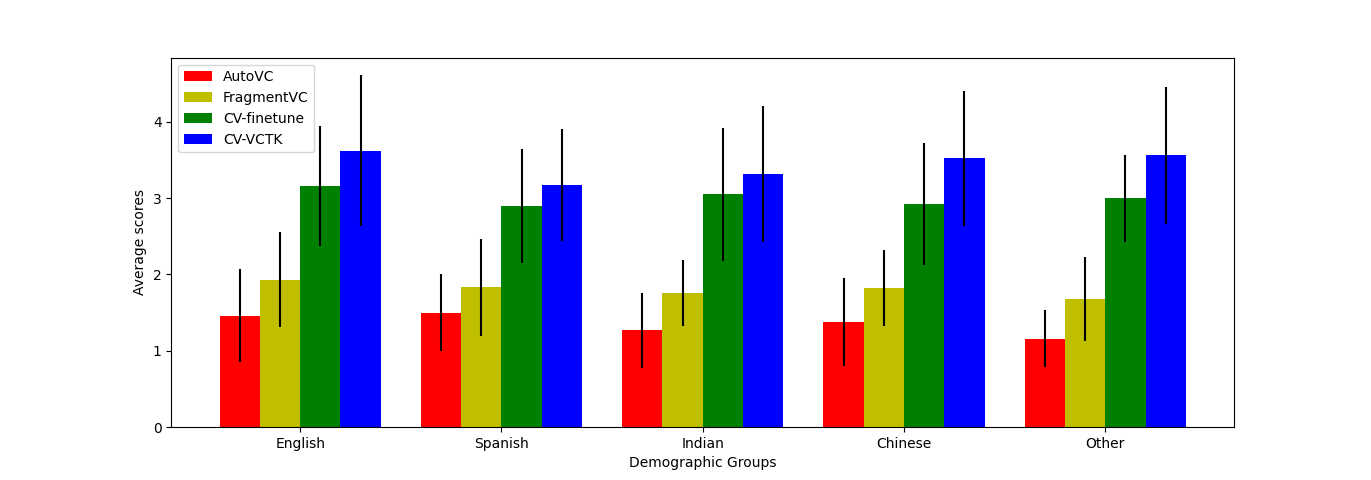}
         \caption{Speakers grouped by demographic groups}
         \label{bar-avg-score-std-demogroups}
     \end{subfigure}
        \caption{Average scores and standard deviations representing model performance of all 4 models across all annotators.}
        \label{fig:three graphs}
\end{figure}

We find that \textbf{yes, this adaptation is easy}. Simply by pretraining on a diverse sample of accented English speech from CommonVoice \cite{ardila2019common} under a variety of input noise conditions,  FragmentVC \cite{lin2021fragmentvc} is able to \textbf{simultaneously convert and denoise diverse speech}, producing clean output speech in the target speaker's voice, while preserving the accent and content of the source utterance.
We open-source our noising data-augmentation scripts and release our noise-robust FragmentVC checkpoint.\footnote{\href{https://github.com/avanitanna/RobustFragmentVC\#checkpoints}{https://github.com/avanitanna/RobustFragmentVC\#checkpoints}} 




\section{Method}
We add a noising module to augment data during training with randomized effects including volume level changes, adding background noise, simulating room reverberation effects and simulating compression artifacts from telephony.These effects are randomly sampled as data is loaded for feature extraction. We train FragmentVC from scratch using a more comprehensive dataset---a combination of CommonVoice and VCTK corpus (checkpoint CV-VCTK)---and perform finetuning of the existing FragmentVC using only CommonVoice (CV-finetune). Both checkpoints are trained for 1 million steps. FragmentVC was optimized with AdamW optimizer and lr 1e-5. 

To assess the robustness of the model, we use a test suite consisting of name clips with multilingual speakers with varying accents. The recordings come from students who self-recorded their full names. The recordings are of varying lengths and differ in audio quality. We ensure that the files represent real-world challenges with audio data (background noise, volume levels, microphone quality, etc.) by further noising them. 
Our code is available on GitHub\footnote{\href{https://github.com/avanitanna/RobustFragmentVC}{https://github.com/avanitanna/RobustFragmentVC}} and the checkpoints are provided there as well\footnotemark[2].

\section{Results}
\begin{table}[ht]
\centering 
\begin{tabular}{c c c c c c} 
\hline\hline 
Annotator & A1 & A2 & A3 & A4 & A5\\ [0.5ex] 
\hline 
A1 & 1 & 0.6 & 0.45 & 0.56 & 0.59 \\ 
A2 & 0.6 & 1 & 0.51 & 0.41 & 0.7 \\
A3 & 0.45 & 0.51 & 1 & 0.3 & 0.45 \\
A4 & 0.56 & 0.41 & 0.3 & 1 & 0.38 \\
A5 & 0.59 & 0.7 & 0.45 & 0.38 & 1 \\ [1ex] 
\hline 
\end{tabular}
\caption{Inter-rater agreement based on average Pearson Correlation (PCC) scores across all models.} 
\end{table}
We present opinion scores from 5 human annotators. In this test, each subject was asked to listen to converted utterances (with the same target speaker) of 64 testing pairs for each of the 4 models - AutoVC, vanilla FragmentVC, CV-finetune FragmentVC, and CV-VCTK-scratch FragmentVC. The subjects were asked to score the models (1 to 5) based on their confidence in how the utterances sounded (1: doesn't sound like speech; 2: sounds like speech, weird noise and incomprehensible; 3: some comprehensible bits, can't fully parse name; 4: can hear what the name is, still some noise or quality issues; 5: clearly can hear the name, sounds clean). The test suite consists of real student name utterances; we do not release the test suite for privacy reasons. Figure 1 shows Likert score histograms for baselines and our robust models. Figure 2 shows the average scores and standard deviation (rounded to 2 decimal places) representing how well the each model performs across all annotators for all source speakers, source speakers grouped by gender (Male or Female voice), and source speakers grouped by self-reported first-name demographic group affiliation (English, Spanish, Indian, Chinese, or Other). We can clearly see that CV-VCTK has better scores than the rest, followed by CV-finetune. We report inter-rater agreement scores in Table 1. 


\section{Conclusion}
We demonstrate denoising capabilities in FragmentVC, providing a denoising objective is used at train time. We address difficulties faced by users in replicating these models and their performance with their own data. We consider real-world challenges with audio data and train FragmentVC to better adapt to these challenges as well as accent variations. We make a more useful checkpoint \footnotemark[2] for real-world, arbitrary, multi-lingual users which can enable them to use such VC models more easily with real-world data. 




\bibliographystyle{IEEEtran}
\bibliography{mybib}

\end{document}